# Evidence for distortion-induced local electric polarization in α-$RuCl_3$

Xinrun Mi [1††], De Hou [2,3††], Xiao Wang [4††], Sabreen Hammouda[4], Caixing Liu [2], Zijian Xiong [1], Han Li [5], Yan Liu [6], Guiwen Wang [6], Aifeng Wang [1], Yisheng Chai [1], Yang Qi [7], Wei Li [8], Xiaoyuan Zhou [1], Yixi Su [4], D. I. Khomskii [9], Mingquan He [1*], Zhigao Sheng [2*], and Young Sun [1*]

[1]College of Physics & Center of Quantum Materials and Devices, Chongqing University, Chongqing 401331, China
[2]Anhui Key Laboratory of Condensed Matter Physics at Extreme Conditions, High Magnetic Field Laboratory, HFIPS, Anhui, Chinese Academy of Sciences, Shushanhu Road 350, Hefei, 230031 China
[3]University of Science and Technology of China, Hefei 230026, China
[4]Jülich Centre for Neutron Science (JCNS) at Heinz Maier-Leibnitz Zentrum (MLZ), Forschungszentrum Jülich GmbH, Lichtenbergstr. 1, D-85747 Garching, Germany
[5]School of Physics, Beihang University, Beijing 100191, China
[6]Analytical and Testing Center, Chongqing University, Chongqing 401331, China
[7]State Key Laboratory of Surface Physics & Department of Physics, Fudan University, 200433, Shanghai, China
[8]CAS Key Laboratory of Theoretical Physics, Institute of Theoretical Physics, Chinese Academy of Sciences, Beijing 100190, China
[9]II. Physikalisches Institut, Universität zu Köln, D-50937 Köln, Germany

*Received January ***; accepted January ***; published online January ****

The spin-orbit assisted Mott insulator α-$RuCl_3$ is a prime candidate for material realization of the Kitaev quantum spin liquid. While little attention has been paid to charge degrees of freedom, charge effects, such as electric polarization, may arise in this system. Here, we report distortion-induced local electric polarization in α-$RuCl_3$ as evidenced by single-crystal X-ray diffraction, second harmonic generation (SHG) and dielectric measurements. The SHG signal appears at room temperature and develops substantially in the Kitaev paramagnetic state when short-range spin correlations come into play. Despite sizable pyroelectric currents in the Kitaev paramagnetic state, the absence of hysteresis in electric field-dependent polarization (*P*-*E*) points to the short-range nature of electric polarization. This localized electric polarization is likely the result of distortion-induced charge dimerization, achieved through virtual hopping-induced charge redistribution. In addition, the electric polarization is boosted by short-range spin correlations via spin-phonon coupling in the Kitaev paramagnetic state. Our results emphasize the importance of charge degrees of freedom in α-$RuCl_3$, which establish a novel platform to investigate charge effects in Kitaev materials.

## 1 Introduction

The intricate interplay of spin, charge, orbital, and lattice degrees of freedom in Mott insulators often gives rise to a broad variety of quantum phases of matter, such as high-temperature superconductivity, frustrated magnetism, and nontrivial topological orders [1, 2]. Of particular interests are spin-orbit assisted Mott insulators, including α-$RuCl_3$, $A_2IrO_3$($A$=Li, Na), which represent prime material playgrounds to pursue the long-sought Kitaev quantum spin liquid (QSL) [3-5]. The Kitaev QSL is characterized by frustrated bond-dependent Kitaev interactions, which can host emergent $\mathbb{Z}_2$-fluxes and Majorana fermion excitations that could be promising ingredients towards topological quantum computing [6, 7]. Exploration of QSL excitations in Mott insulating Kitaev materials, both in magnetic and charge channels, is of great fundamental and technical importance[8, 9].

*Corresponding author (email: mingquan.he@cqu.edu.cn; zhigaosheng@hmfl.ac.cn; youngsun@cqu.edu.cn)
††These authors contributed equally to this work.

The Mott insulator α-$RuCl_3$ is in close proximity to the Kitaev QSL [10-16]. Edge-sharing $RuCl_6$ octahedra form a layered structure stacking along the *c*-axis and Ru atoms are coordinated in a honeycomb lattice within the *ab*-plane [Figure 1(a)]. The effective interactions among nearest neighboring Ru spins can be described by the bond-directional *S*-1/2 Kitaev model [10, 11]. Although finite non-Kitaev interactions drive α-$RuCl_3$ into a zigzag antiferromagnetic (AFM) phase below $T_N$ ~ 7 K, Kitaev fluctuations and fractional excitations have been found to survive up to $T_H$ ~ 70-100 K in the Kitaev paramagnetic state [14-16]. Moreover, the QSL state may arise when the AFM order is suppressed by an in-plane magnetic field above ~ 8 T [12, 13].

Compared with intensive studies in the spin channel, charge degrees of freedom are often overlooked in Kitaev materials due to strong on-site Coulomb repulsion. Nevertheless, electrons are allowed to virtually visit neighboring sites, which actually is the key ingredient to mediate exchange interactions in the spin channel. Theoretically, such virtual hopping of electrons can give rise to various intriguing charge effects, such as fractional charges, spontaneous orbital currents, and electric polarization [17-20]. This scenario is likely realized in the spin-orbit entangled Kitaev material α-$RuCl_3$ [21, 22], which has sizable electronic dipole contributions to the observed sub-gap optical conductivity [23-26]. α-$RuCl_3$ has moderate electronic correlations ($U$ ~ 2 eV) and sizable nearest-neighbor hopping ($t$ ~ 100 meV) [21, 22]. Finite values of $t/U$ and wide-spread *d*-orbitals are advantageous for the development of charge fluctuations. Even more interesting in this respect is the possibility to access, and even manipulate low-energy QSL excitations electrically via these charge effects [8, 9, 27].

In this article, we have discovered local electric polarization in α-$RuCl_3$. Virtual hopping induced charge fluctuations and moderate in-plane distortions are likely the key ingredients to promote local electric polarization, which gets enhanced in the Kitaev paramagnetic state via magnetoelastic coupling. Our findings suggest that α-$RuCl_3$ is a prime system to investigate charge effects in Kitaev materials. The enhanced electric polarization in the Kitaev paramagnetic state in α-$RuCl_3$ suggests the intertwined charge and spin channels, which could pave the way for electrical control of QSL excitations.

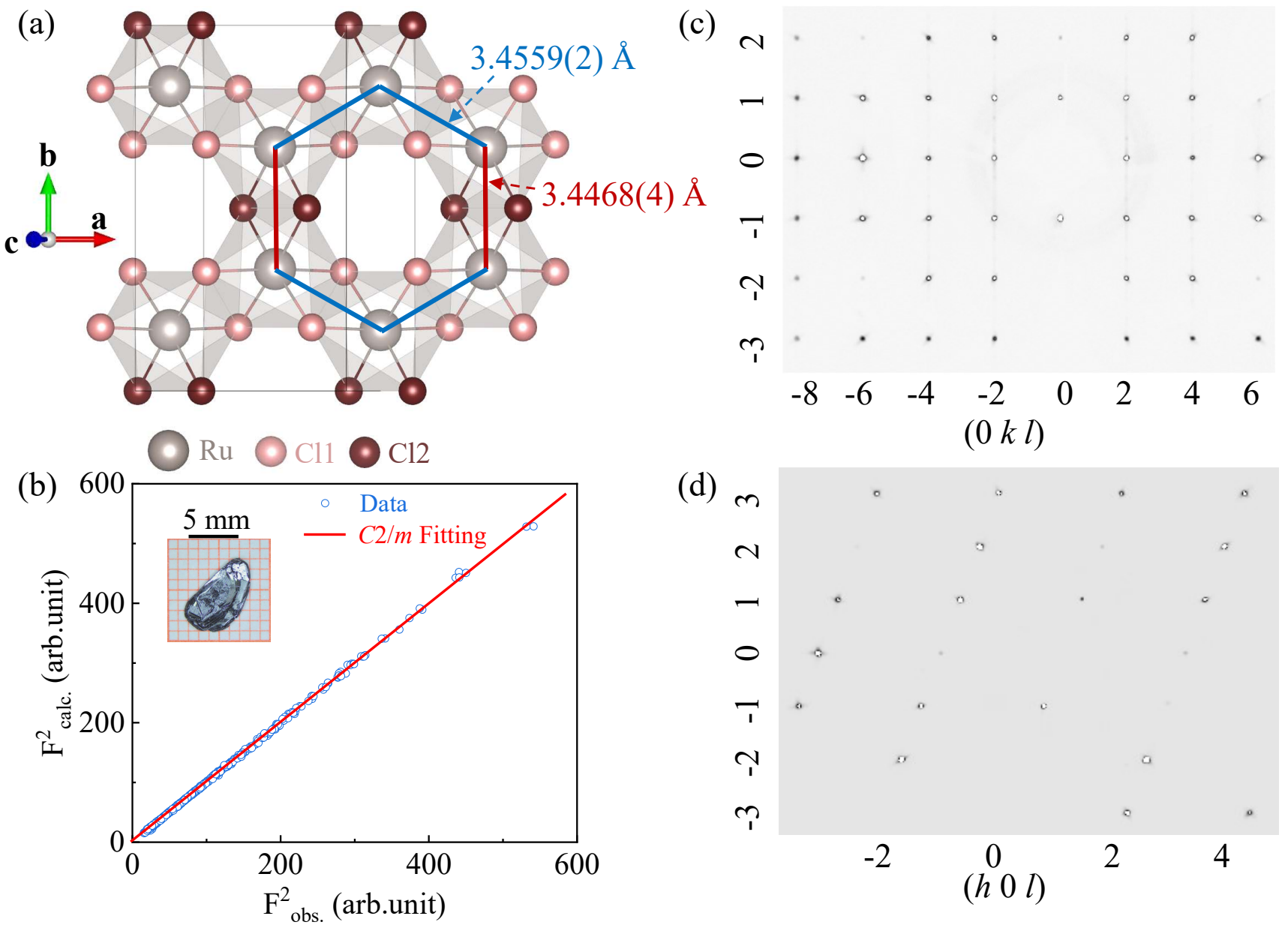

**Figure 1** Room-temperature crystal structure of α-$RuCl_3$ determined by single-crystal XRD. (a) Ru atoms form a honeycomb structure within the *ab*-plane. The honeycomb lattice is slightly distorted with two types of Ru-Ru bond length (red: 3.4468(4) Å, blue: 3.4559(2) Å). (b) The observed squared structure factors can be well fitted with the monoclinic *C*2/*m* structure. (c) and (d) Single-crystal XRD patterns obtained for (0*kl*) and (*h*0*l*) planes, respectively.

**Table.1** Refined *C*2/*m* structure parameters of α-$RuCl_3$ obtained from single-crystal XRD experiments performed at room temperature. The cell parameters are: $a$ = 5.98440(10) Å, $b$ = 10.3519(3) Å, $c$ = 6.0452(2) Å, $\alpha = \gamma$ = 90°, $\beta$ = 108.753(2) ° .

| Atom | Site | $x$ | $y$ | $z$ | $U_{iso}$ | $U_{11}$ | $U_{22}$ | $U_{33}$ | $U_{12}$ | $U_{13}$ | $U_{23}$ |
|---|---|---|---|---|---|---|---|---|---|---|---|
| Ru | 4h | 0.5 | 0.16648(18) | 0.5 | 0.01224(9) | 0.01154(15) | 0.01068(14) | 0.01493(14) | 0 | 0.00484(9) | 0 |
| Cl1 | 4i | 0.27332(11) | 0 | 0.26514(11) | 0.01720(18) | 0.0162(3) | 0.0128(3) | 0.0198(3) | 0 | 0.0018(2) | 0 |
| Cl2 | 8j | 0.25067(9) | 0.32583(3) | 0.26772(8) | 0.01703(14) | 0.0167(2) | 0.0171(2) | 0.0191(2) | 0.0191(2) | 0.00830(17) | 0.00404(13) |

## 2 Materials and method

*Crystal growth*

High-quality single crystalline α-$RuCl_3$ samples were grown in a two-zone furnace using the chemical vapor transport method [28]. Commercial $RuCl_3$ powder (Furuya metal) was firstly sealed in a silica ampule, which was subsequently inserted into a two-zone furnace. The source and sink temperatures were set to 790 °C and 710 °C, respectively. Single crystals of α-$RuCl_3$ were obtained as black shiny plates at the sink end after 5 days.

*Single crystal X-ray diffraction*

In-house single-crystal X-ray diffraction (XRD) was performed at room temperature using a Rigaku XtaLAB Synergy-S diffractometer with Mo-$K_\alpha$ (0.709 Å). The CrysAlisPro software was used for the data reduction and Jana2006 for the structural refinement. Total 2091 reflections used in the refinement merged into 475 reflections.

*Pyroelectric, dielectric and polarization-electric field loop measurements*

Two typical types of samples were chosen for pyroelectric measurements. Sample 1ab and sample 4ab (see Supplementary Information Figure S3) are relatively thick crystals with typical dimension of (5×3×0.6 $mm^3$). These two samples were cut into rectangular cuboids with the largest faces perpendicular to the honeycomb layer, so that an $\mathbf{E} \parallel ab$ poling configuration was adopted. The other type of samples, i.e., sample 2c, sample 3c, sample 5c and sample 6c, are thin plates with the widest planes being the *ab* plane (typical dimension: 5×5×0.05 $mm^3$). And an $\mathbf{E} \perp ab$ poling geometry was used for the second type of samples. Thin layer of Au (50 nm) was sputtered on the corresponding sample surfaces to serve as contact electrodes for the pyroelectric measurements. Background current was measured by a Keithley 6517B electrometer upon heating (3 K $min^{-1}$) after cooling the samples down to 2 K in zero electric field. Subsequently, poling electric fields were applied at 80 K using the Keithley 6517B and the samples were then cooled down in electric fields down to 2 K. The pyroelectric current was finally recorded using the Keithley 6517B upon warming (3 K $min^{-1}$) in zero electric field after the electrodes were shortened for 20 min at 2 K. The net pyroelectric current was evaluated by subtracting the background current from the measured total pyroelectric current. The pyroelectric measurements were carried out in a commercial cryostat (Oxford Instruments, 14 T).

The dielectric properties of α-$RuCl_3$ samples were measured using an Agilent E4980A LCR meter. The polarization-electric field (*P-E*) loops were characterized by a ferroelectric analyzer (TF Analyzer 2000E).

*Second-harmonic generation (SHG)*

The SHG measurements were conducted in a low-vibration optical cryostat (ARS, CS204SF-FMX-20) with a base temperature of 10 K. Femtosecond fundamental beams (800 nm) were generated from a Ti: sapphire laser (100 fs, 84 MHz). A normal incidence geometry was adopted, and the probing electric fields of light were directed within the *ab* plane of α-$RuCl_3$ samples. The SHG signal was collected using two configurations, XX and YY, in which the fundamental light and the outgoing beam were co- and cross-polarized, respectively. Samples were cleaved in-situ before SHG experiments.

*Heat capacity experiment*

The specific heat $C_P$ experiments were performed in a Physical Property Measurement System (Quantum Design Dynacool, 9 T) using the conventional relaxation method from 2 K to 200 K upon warming. The lattice contribution was estimated by the specific heat of a non-magnetic isostructural compound $RhCl_3$ using the data reported by S. Widmann et al [16] (see Supplementary Information for more details).

## 3 Results and discussion

The crystal structure of α-$RuCl_3$ has been the subject of intense debate. Depending on the inter-layer stacking details, a variety of crystal structures have been reported at room temperature, including trigonal $P3_112$ [11, 29, 30], monoclinic $C2/m$ [31-33], and rhombohedral $R\overline{3}$ [34, 35]. A switching from the high-temperature $C2/m$ polytype to the low-temperature $R\overline{3}$ phase [34], or from $P3_112$ to $C2/m$ [30], at a highly hysteric first-order structure transition ($T_S \sim 150$ K) has been suggested. It seems that the room-temperature structure is widely accepted as $C2/m$, while the low-temperature structure remains controversial [36]. Figure 1 shows the single-crystal XRD results obtained at room temperature for a typical α-$RuCl_3$ sample used in this study. The observed sharp Bragg peaks can be well described by the $C2/m$ representation, which follows the *C*-centering selection rule of $h + k = 2n$ ($n$ = integer). The $2/m$ profile can be exemplified by identifying the twofold rotational axis $\mathbf{b^*}$ and the mirror-plane perpendicular to $\mathbf{b^*}$ [see Figure 1(c)]. The refined room-temperature structure parameters are listed in Table I. In addition to the sharp Bragg peaks, weak diffuse scattering spots are also seen at $l = n \pm 1/3$, indicating the existence of minor stacking faults [31, 32]. Moreover, as shown in Figure 1(a), the honeycomb lattice is slightly distorted in a way that one type of Ru-Ru bonds [red, 3.4468(4) Å] are shorter than other two types (blue, 3.4559(2) Å) by ~ 0.2%, agreeing well with earlier reports [24, 30, 32]. This in-plane distortion may turn out to be essential for the establishment of electric polarization in such a bipartite lattice, as we will discuss later.

To investigate the charge effects, we start with pyroelectric experiments, and the results are presented in Figure 2. Two typical single crystals, sample 1ab and sample 2c, were studied with poling electric fields applied in-plane (**E** ∥ *ab*) and out-of-plane (**E** ∥ *c*), respectively. Similar results are found in other samples (Supplementary Information Figure S2 and S3). To avoid the impacts of the hysteric structure transition around 150 K, all measurements in this study were performed upon warming. A broad peak centered around $T_H \sim 60$ K is seen in both samples, as shown in Figure 2(a) and 2(b). In addition, the sign of the pyroelectric current is switchable by reversing the direction of poling electric fields. At first glance, these findings may suggest the existence of ferroelectricity in α-$RuCl_3$. However, as seen in Figure 2(c) and 2(d), the electric polarization ($P$) varies linearly with the electric field applied both in-plane and out-of-plane. No clear signatures of hysteresis are found in the *P-E* loop, ruling out the appearance of long-range ferroelectricity. It appears that extrinsic effects, such as trapped charges, contribute significantly to the observed pyroelectric current, although the pyroelectric current peaks at the same temperature when measured at different heating rates (Supplementary Information Figure S3).

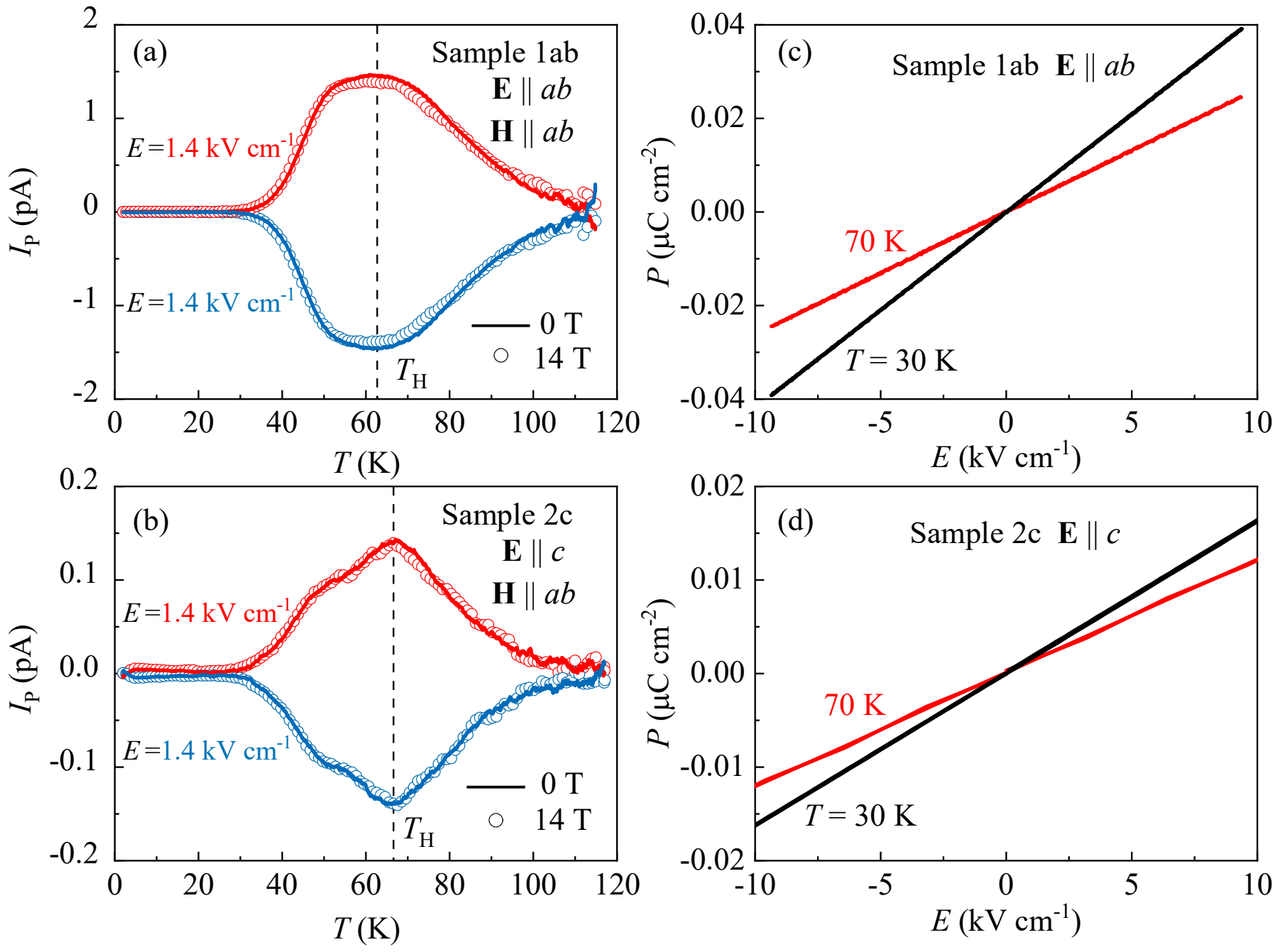

**Figure 2** Pyroelectric current and electric polarization. (a) and (b) The pyroelectric current ($I_P$) of sample 1ab (**E** ∥ *ab*) and sample 2c (**E** ∥ *c*), respectively. For both samples, $I_P$ peaks around $T_H \sim 60$ K (vertical dash lines), which responds weakly to in-plane magnetic fields up to 14 T (open symbols). (c) and (d) The directly measured electric polarization ($P$) as a function of electric field recorded at fixed temperatures for the corresponding samples.

Figure 3 shows the temperature-dependent dielectric constant $\varepsilon'$ and conductivity $\sigma' = \omega\varepsilon''\varepsilon_0$ measured for sample 2c and sample 5c, using electric fields applied out-of-plane (**E** || *c*). Similar results are found in sample 1ab with **E** || *ab* [see Figure S1]. Here, $\omega = 2\pi f$ is the angular frequency, $\varepsilon''$ is the dielectric loss, $\varepsilon_0$ represents the dielectric constant of vacuum. The relative dielectric constant is defined as $\varepsilon' = Cd/\varepsilon_0 A$, where $d$ is the sample thickness, $C$ is the measured total capacitance, and $A$ is the contact area of the electrodes. The dielectric constant is highly frequency-dependent at temperatures above $T_H \sim 60$ K, below which negligible frequency response is found. Near 100 K, a bump appears in $\sigma'$ at a temperature $T_b$, which moves to higher values with increasing frequency [see Figure 3 (b)]. Around 200 K, a peak is observed in $\varepsilon'$ in both samples. This peak also shifts gradually towards higher temperatures with increasing frequency. The peak in $\varepsilon'$ likely has an Maxwell-Wagner origin, which arises from electrode polarization due to the formation of Schottky depletion layers at the electrode-sample interfaces [37]. To study the impacts of interfacial effects further, we compare the dielectric constant measured for sample 5c using different contacts in Figure 3(c) and 3(d). At high temperatures above 100 K, $\varepsilon'$ obtained for sputtered gold contacts is much larger than that of silver-paint contacts, especially at low frequencies. This implies that the high-temperature dielectric constant is dominated by interfacial contributions. However, below 100 K, the differences in $\varepsilon'$ measured using sputtered contacts and silver-paint are less evident. As seen in Figure 3(d), the bump in $\sigma'$ around 100 K is observed using both types of contacts. A hump structure is also seen at $T_b$ in $\varepsilon'$ upon closer inspections [see Figure 3(e)]. As displayed in Figure 3(f), the frequency-dependent $T_b$ can be described well by the Vogel-Fulcher law $f = f_0[-E_a/\kappa_B(T_b - T_f)]$, where $f_0$ is the Debye frequency, $E_a$ is the activation energy, $\kappa_B$ is the Boltzmann's constant, $T_f$ is the freezing temperature. This kind of relaxation near 100 K is very likely caused by the freezing of short-range electric polarization.

The appearance of short-range electric polarization is further supported by the observation of sizable SHG signal as shown in Figure 4. The SHG results were obtained for sample 2ab ($\mathbf{E}_{\omega,2\omega}$ ∥ *ab*) using a normal incidence geometry, which was not poled by an external electric field. Here, we focus on a linearly co-polarized configuration (XX) of fundamental and second harmonic beams. Data collected using a linearly cross-polarized (XY) geometry can be found in Supplementary Information Figure S5. As shown in Figure 4(a), sizable SHG signal (red spheres) is found below $T_H$, which persists up to room temperature with significantly reduced intensity. The SHG signal only varies slightly at low temperatures below $T_H$, whereas further warming leads to rapid loss of intensity. Negligible variations are again recovered above 100 K. In the crossover region (50 K ~ 90 K), we note that error bars of the SHG signal are significantly enhanced, compared with other temperatures. In addition, the illuminated spot of the sample surface became greenish when passing through the crossover temperature interval (see Supplementary Information Figure S6), indicating sizable changes in dielectric properties. These effects again point to the development of fluctuating, short-range electric dipole-dipole correlations below 100 K.

Figure 4(c) and 4(d) present the polar plots of SHG signal. Here, the polarization directions of fundamental and excited detection beams were rotated simultaneously within the sample *ab* plane. The $\phi = 0$ direction is perpendicular to the Ru-Ru bounds. At first glance, the SHG data captured at 300 K resembles a threefold rotational symmetry. However, deviations from the ideal $C_3$ symmetry can be identified upon closer inspections. The SHG signal can be more appropriately characterized by a $C_2$ symmetry, in accordance with the $C2/m$ crystal structure and distorted honeycomb layer as determined by the single-crystal XRD experiments [see Figure 1]. At low temperatures below 100 K, the $C_2$ symmetry becomes more evident. More plots captured at other temperatures are shown in Supplementary Information Figure S5. Similar $C_2$ symmetric properties in the low-temperature phase have also been documented in prior studies, including anisotropic in-plane susceptibility [38], magnetotropic coefficient [39] and specific heat measurements [40]. As illustrated in Figure 4(c) and 4(d), the SHG signal, measured at both 300 and 10 K, can be effectively described by a $C_2$ point group (solid lines) with a twofold axis directed along the *b*-axis. The persistence of $C_2$ symmetry, from room temperature down to low temperatures, has also been observed in earlier Terahertz and single-crystal XRD reports [24, 32, 41].

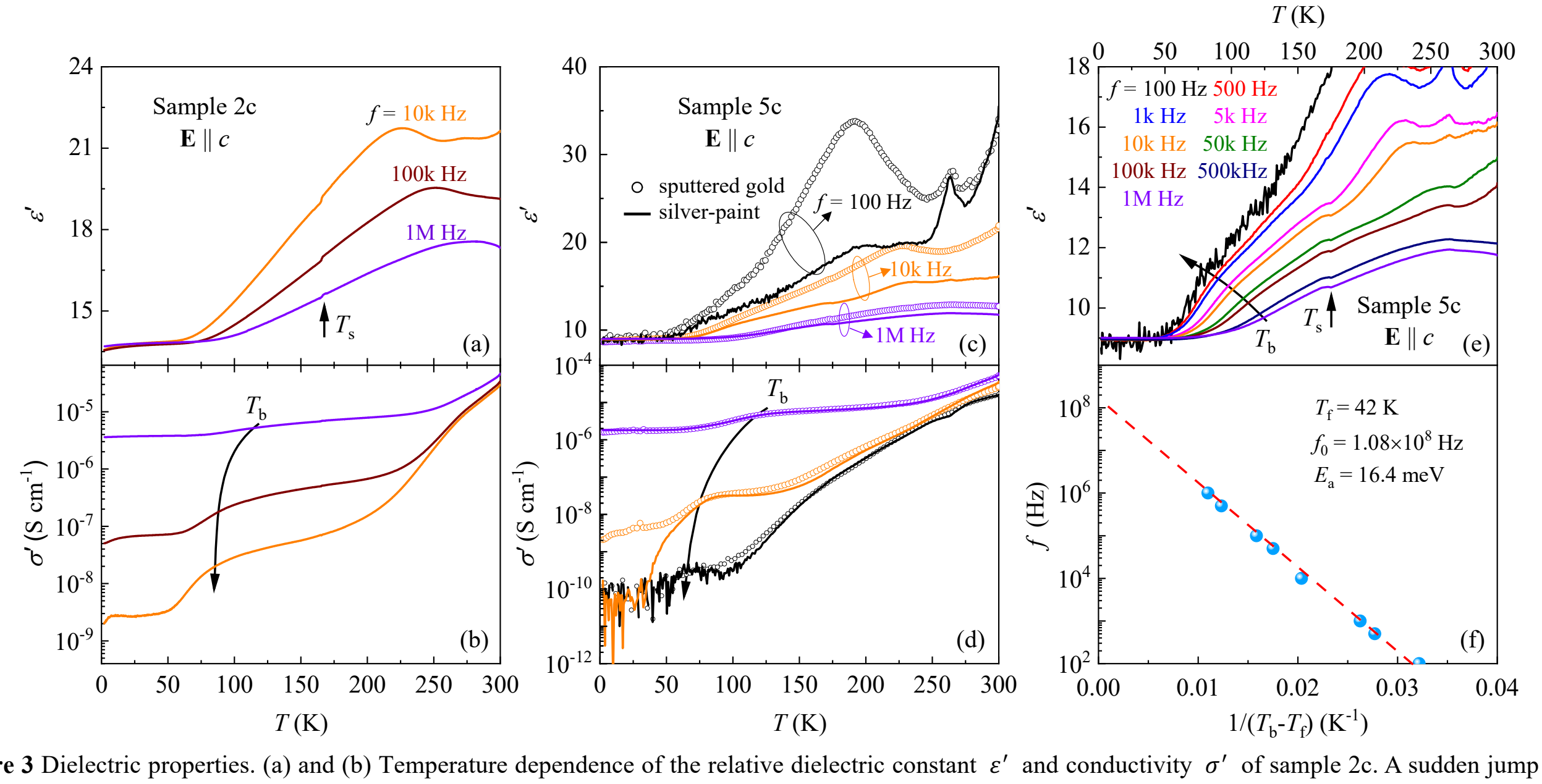

**Figure 3** Dielectric properties. (a) and (b) Temperature dependence of the relative dielectric constant $\varepsilon'$ and conductivity $\sigma'$ of sample 2c. A sudden jump around $T_s$ ~ 170 K is seen in $\varepsilon'$, which is a signature of the structure transition. (c) and (d) Comparison of the dielectric constant and conductivity measured for sample 5c using sputtered gold (open symbols) and silver-paint (solid lines) contacts. (e) Enlarged view of $\varepsilon'$ obtained for silver-paint contacts. (f) Fitting of the frequency-dependent $T_b$ (solid symbols) according to the Vogel-Fulcher law (dashed line).

To further unveil the origin of the enhanced SHG intensity below 100 K, we compare the SHG signal with the excess specific heat in Figure 4(a). The excess specific heat $C_{ex} = C_{total} - C_{ph}$ is obtained by subtracting the phonon contribution $C_{ph}$ from the measured total specific heat $C_{total}$. The phonon background is estimated from an isostructural non-magnetic material $RhCl_3$ following S. Widmann et al [16](see Supplementary Information Figure S4 for more details). A sharp peak is seen in $C_{ex}$ at $T_s$ = 170 K, which is caused by the first-order structure transition. At lower temperature around $T_H$, a broad hump appears in $C_{ex}$, which has been attributed to the entropy release of Majorana fermion excitations in the Kitaev framework [15, 16, 42, 43]. Signatures of fractional excitations emerging near $T_H$ have also been evidenced by various other techniques [10-16]. It is well established that short-range spin correlations start to develop in the vicinity of $T_H$, regardless of their microscopic origin. Interestingly, the SHG also changes most rapidly around $T_H$. The evolution of short-range electric polarization certainly is correlated with the establishment of short-range spin correlations. In addition to the high-temperature peak, there are two other anomalies appearing in specific heat at $T_{N1}$ ~ 7 K and $T_{N2}$ ~ 14 K, which are signatures of long-range ordered zigzag AFM transitions. Appearance of multiple transitions suggests the existence of stacking faults. It has been suggested that an *ABC* inter-layer stacking gives rise to the transition at $T_{N1}$, whereas an *ABAB* stacking leads to the transition at $T_{N2}$ [31]. Formation of stacking faults can break the inversion symmetry locally at *ABAB* and *ABC* stacking interfaces, leading to local electric polarization [28]. The degrees of stacking faults are expected to increase considerably below the structure transition [32, 41]. However, the SHG intensity only changes moderately below and above $T_s$, suggesting minor contributions from staking faults. In addition, the structure transition primarily influences interlayer stacking. The lack of correlation between the SHG signal and both the interlayer stacking and the degree of stacking faults supports the hypothesis that in-plane lattice distortion is responsible for the observed SHG signal.

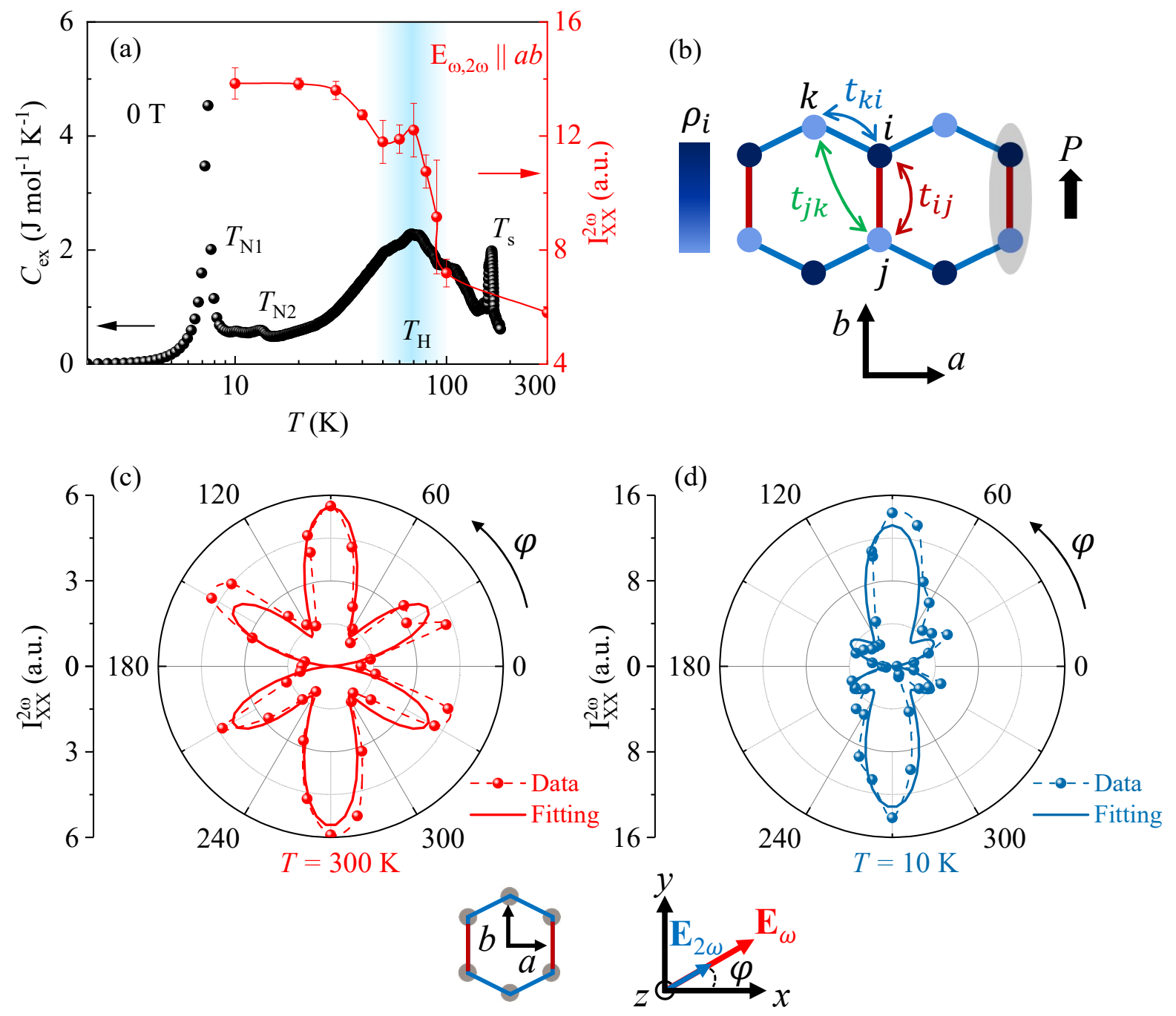

**Figure 4** Second harmonic generation (SHG). (a) Temperature dependence of SHG intensity ($I^{2\omega}_{XX}$, red spheres) measured for sample 2ab using a normal incidence geometry ($\mathbf{E}_{\omega,2\omega}$ || *ab*). The fundamental and second harmonic beams were linearly co-polarized. Excess specific heat (black spheres) is also presented for comparison. (b) Schematic illustration of charge modulation induced local electric polarization in α-$RuCl_3$. Charge density ($\rho_i$) distributes inhomogeneously on a distorted lattice due to anisotropic hopping ($t_{ij} \neq t_{ki}$) along nearest neighboring bonds, forming a local trigonal charge order and charge dimerization (shaded oval), which gives rise to electric polarization along the *b* direction. (c) and (d) Polar plots of SHG intensity recorded at 300 K and 10 K. The sample lab coordinates are denoted as (*x*, *y*, *z*) and the $\varphi$ = 0 direction points perpendicular to the Ru-Ru bounds. Solid lines are theoretical fitting the $C_2$ symmetry. Dashed lines are guide to the eye.

As indicated by the XRD results in Figure 1, the room-temperature structure is *C*2/*m*, which is centrosymmetric and cannot account for the substantial SHG signal observed at room temperature. The finite SHG response may potentially stem from electric-quadrupole interactions within centrosymmetric structures. However, no detectable electric-quadrupole interactions were observed in early $^{99}$Ru Mössbauer spectroscopy [44]. Another possible driving force of the local electric polarization could be distortion-induced charge dimerization, as also found in molecular dimer Mott insulators [45, 46]. Indeed, short-range trigonal and dimer-like charge orders have been visualized in α-$RuCl_3$ by scanning tunneling microscopy (STM) [30]. For the $4d^5$ configuration of $Ru^{+3}$ valence state, one hole per site is expected without any polarization. However, the average charge density per site [$\rho_i = e(1+\delta\rho_i)$] can deviate from unit charge (*e*) due to virtual charge hopping among neighboring sites [18, 19]. As shown in Figure 4(b), to mimic the case of α-$RuCl_3$, we consider a distorted two-dimensional honeycomb lattice, i.e., bonds along the *b* direction differ slightly from other two types. The electron density operator $n_i$ at a given site *i* is related to spin-spin correlations [18, 19]:

$$n_i = 1 + \sum_{jk}\left[\frac{8t_{ij}t_{jk}t_{ki}}{U^3}(\mathbf{S}_i\cdot\mathbf{S}_j + \mathbf{S}_i\cdot\mathbf{S}_k - 2\mathbf{S}_j\cdot\mathbf{S}_k)\right], \quad (1)$$

where the sum runs over the shortest closed triangular (*ijk*) loops connected with the site $i$, $t_{ij}$ is the hopping integral between sites $i$ and $j$, $\mathbf{S}_i$ is the spin operator at site $i$. The virtual hopping induced charge imbalance ($\delta\rho_i$) at site $i$ is proportional to the spin correlation function:

$$\delta\rho_i = A\sum_{jk}\langle\mathbf{S}_i\cdot\mathbf{S}_j + \mathbf{S}_i\cdot\mathbf{S}_k - 2\mathbf{S}_j\cdot\mathbf{S}_k\rangle, \quad (2)$$

where *A* is a constant related to hopping integral and on-site repulsion. Employing the physical parameters relevant to α-$RuCl_3$, R. G. Pereira et al. determined $A \cong 1.9\times10^{-4}$ [9]. The charge imbalance $\delta\rho_i$ is nonzero if the average spin correlation function is nonzero, which can be realized in distorted lattice with anisotropic hopping ($t_{ij} \neq t_{ik}$), particularly in the Kitaev paramagnetic state characterized by short-range spin correlations. This charge imbalance on distorted lattices naturally explains the charge dimerization as observed by STM [30]. Chain-like structures, which break the $C_3$ symmetry of an ideal honeycomb lattice, are also seen by transmission electron microscopy (TEM), pointing to dimerization-induced structural reorganization [Supplementary Information Figure S7]. The inversion symmetry is broken by this kind of site- and bond-centered charge dimerization, as also seen in other charge-order-induced ferroelectrics, such as $Pr_{1-x}Ca_xMnO_3$ and $Fe_3O_4$ [47, 48]. By symmetry, in-plane electric polarization is locally formed along the *b*-axis [see Figure 4(b)]:

$$P \propto \delta\rho_i e d, \quad (3)$$

where $d$ represents the bond length of charge dimers. In the Kitaev paramagnetic state, the emergence of short-range spin correlations strengthens the spin correlation function, leading to increased electric polarization and rapid enhancement of the SHG intensity below $T_H$. Notably, application of in-plane electric fields can moderately tune the magnetization of α-$RuCl_3$ below $T_H$ [Supplementary Information Figure S8], further underscoring the intimate coupling between charge and spin degrees of freedom. Additionally, magnetoelastic coupling may contribute to lattice distortions, which in turn result in elevated electric polarization and an intensified SHG signal. Indeed, sizable magnetoelastic coupling has been revealed by substantial phonon softening and hardening effects below ~ 100 K and ~ 70 K for different phonon modes, as found in inelastic X-ray and Raman scattering experiments [49, 50]. Enhanced SHG intensity resulting from short-range spin correlations through magnetoelastic coupling has also been observed in other van der Waals magnets [51, 52].

Local charge imbalance induced polarization is also possible even in the pure Kitaev scenario [9, 21, 22], which likely explains the sub-gap optical conductivity found by terahertz spectroscopy experiments [23-26]. There, the electric polarization arises at the second order of hopping amplitude in the presence of finite trigonal distortion, Hund's coupling, and strong spin-orbit coupling. Moreover, as shown by R. G. Pereira et al [9], finite charge modulations and electric polarization can arise around $\mathbb{Z}_2$ vortices, allowing electrical control of gauge fluxes and fractional spin excitations [9]. However, one may expect clear cross-tuning effects of electric (magnetic) properties using magnetic (electric) fields, which are not evident in our experiments. Nevertheless, this possibility cannot be completely ruled out, since extremely large electric fields (E ~ $10^7$ V $m^{-1}$) and sufficiently low temperatures (≪1 K) may be required to see sizable cross-tuning effects [27]. Further studies are necessary to extract the interplay of electric polarization and low-energy spin excitations.

## 4 Conclusions

In summary, local electric polarization has been found in α-$RuCl_3$. The electric polarization is attributed to charge dimerization resulting from lattice distortions and virtual hopping-induced charge redistributions. Furthermore, short-range spin correlations in the Kitaev paramagnetic state enhance electric polarization through magnetoelastic coupling. Our findings highlight α-$RuCl_3$ as a unique system to investigate charge effects in Kitaev materials, and call for further investigations into the interplay of spin and charge channels.

*We thank Long Zhang, Yuan Li and Xuefeng Zhang for stimulating discussions. A portion of this work was performed at the Anhui Laboratory of High Magnetic Field with the help of Yuyan Han. We thank Dr. Bin Zhang at the Analytical and Testing Center of Chongqing University for the assistance on TEM experiments. This work has been supported by National Natural Science Foundation of China (Grant Nos. 12047564, 51725104), Chinesisch-Deutsche Mobilitätsprogamm of Chinesisch-Deutsche Zentrum für Wissenschaftsförderung (Grant No. M-0496), the Open Fund of the China Spallation Neutron Source Songshan Lake Science City (Grant No. DG2431351H). The work of D. I. Khomskii was funded by the Deutsche Forschungsgemeinschaft (DFG, German Research Foundation) - Project number 277146847 - CRC 1238. Y. Chai acknowledges the support by National Natural Science Foundation of China (Grant No. 12227806). W. Li acknowledges the support by National Natural Science Foundation of China (Grant Nos. 11974036, 11834014).*

## Supplementary materials

More experimental data can be found in Supplementary Information

# Supporting Information for

# Evidence for distortion-induced local electric polarization in α-$RuCl_3$

Xinrun Mi [1††], De Hou [2,3††], Xiao Wang [4††], Sabreen Hammouda[4], Caixing Liu [2], Zijian Xiong [1], Han Li [5], Yan Liu [6], Guiwen Wang [6], Aifeng Wang [1], Yisheng Chai [1], Yang Qi [7], Wei Li [8], Xiaoyuan Zhou [1], Yixi Su [4], D. I. Khomskii [9], Mingquan He [1*], Zhigao Sheng [2*], and Young Sun [1*]

[1]*College of Physics & Center of Quantum Materials and Devices, Chongqing University, Chongqing 401331, China*

[2]*Anhui Key Laboratory of Condensed Matter Physics at Extreme Conditions, High Magnetic Field Laboratory, HFIPS, Anhui, Chinese Academy of Sciences, Shushanhu Road 350, Hefei, 230031 China*

[3]*University of Science and Technology of China, Hefei 230026, China*

[4]*Jülich Centre for Neutron Science (JCNS) at Heinz Maier-Leibnitz Zentrum (MLZ), Forschungszentrum Jülich GmbH, Lichtenbergstr. 1, D-85747 Garching, Germany*

[5]*School of Physics, Beihang University, Beijing 100191, China*

[6]*Analytical and Testing Center, Chongqing University, Chongqing 401331, China*

[7]*State Key Laboratory of Surface Physics & Department of Physics, Fudan University, 200433, Shanghai, China*

[8]*CAS Key Laboratory of Theoretical Physics, Institute of Theoretical Physics, Chinese Academy of Sciences, Beijing 100190, China*

[9]*II. Physikalisches Institut, Universität zu Köln, D-50937 Köln, Germany*

[††] These authors contributed equally to this work

*Corresponding author. Email: mingquan.he@cqu.edu.cn; zhigaosheng@hmfl.ac.cn; youngsun@cqu.edu.cn.

**Dielectric Constant**

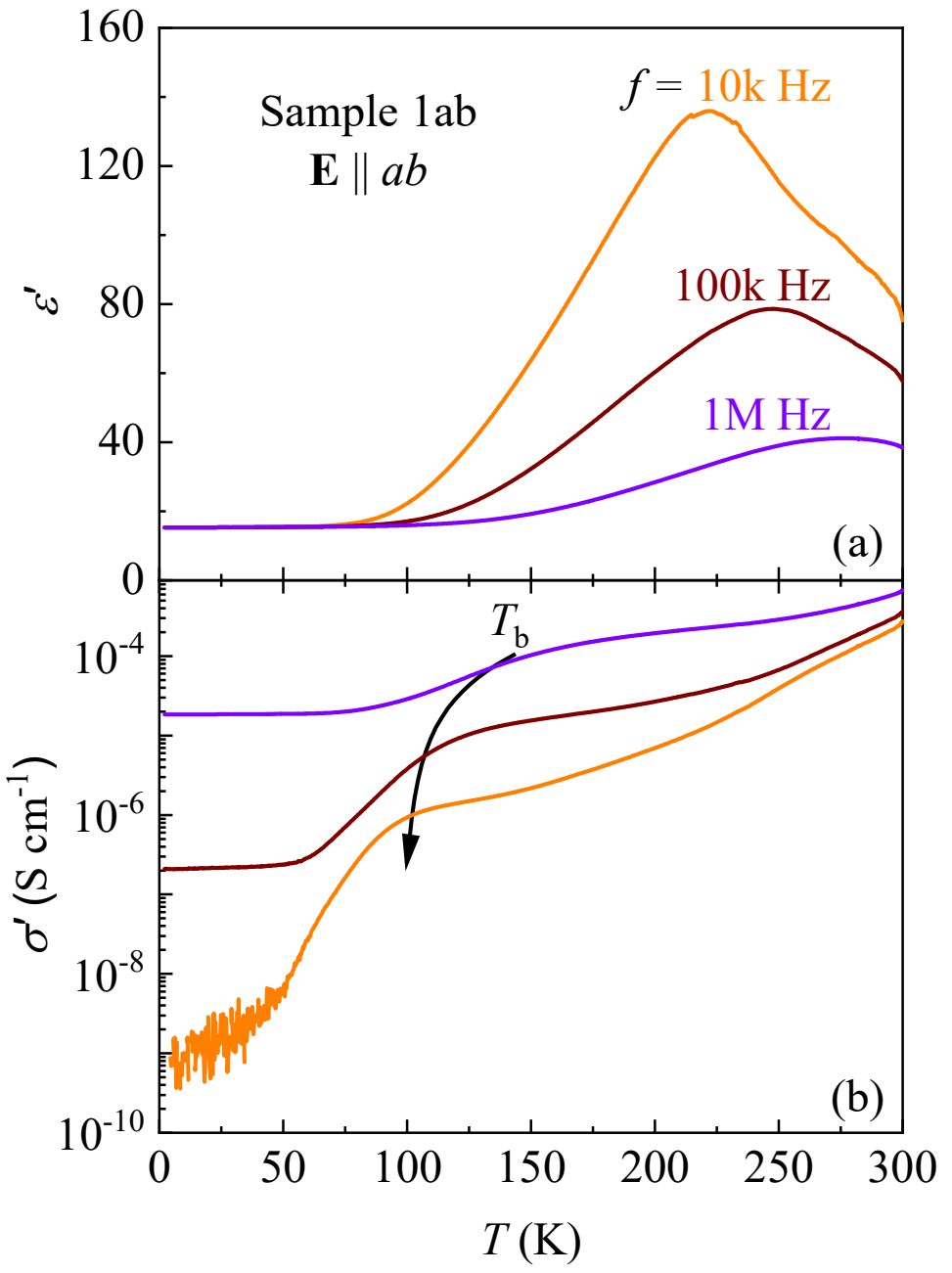

**Figure S1** Temperature dependence of the relative dielectric constant $\varepsilon'$ and conductivity $\sigma'$ of sample 1ab.

**Pyroelectric experiments**

Figure S2 shows the experimental raw data of the pyroelectric measurements. Background currents [black lines in Figure S2(a) and (b)] were firstly measured without poling by an external electric field. The pyroelectric currents were subsequently recorded after poling. The net pyroelectric current shown in the main part of the manuscript was evaluated by subtracting the background current. In Fig. S2(c), the effects of external out-of-plane magnetic field on the pyroelectric current of sample 1ab are presented. Similar to that of in-plane magnetic field, no clear response is found. Figure S2(d) displays the pyroelectric current of another sample (sample 4ab). Similar results are found compared with that of sample 1ab using the same poling configuration (**E** || *ab*).

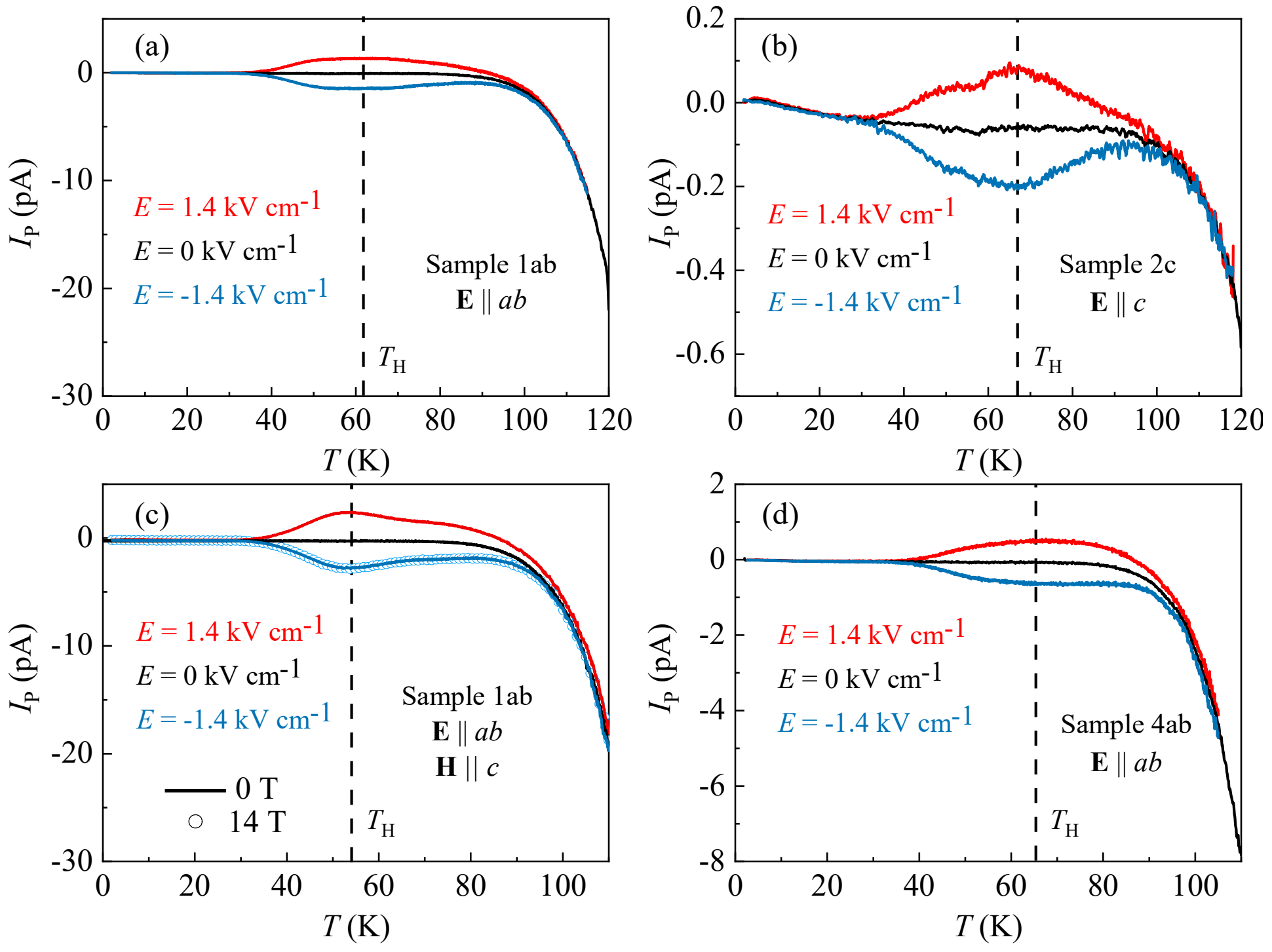

**Figure S2** (a, b) Experimental raw data of the measured pyroelectric current ($I_P$) as a function of temperature for sample 1ab and sample 2c. Black lines are background currents recorded without poling. (c) The pyroelectric current of sample 1ab, which is nearly independent of external magnetic fields applied out-of-plane up to 14 T. (d) The pyroelectric current of sample 4ab measured using an **E** || *ab* poling geometry.

In Figure S3, the pyroelectric currents measured using different heating rates are shown for sample 5c and sample 6c. In both samples, the peak in pyroelectric current stays at the same temperature at different temperature ramping rates.

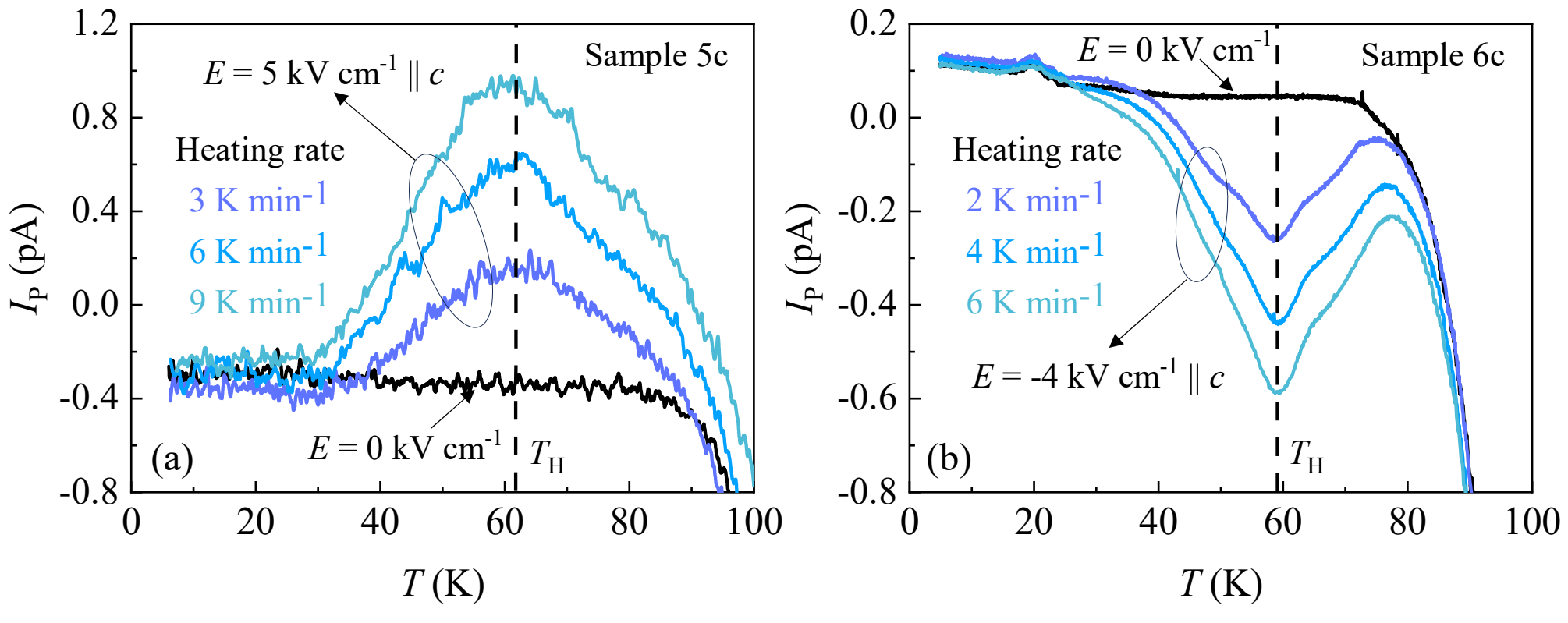

**Figure S3** Pyroelectric current measured for sample 5c and sample 6c using different heating rates. The peak position in pyroelectric current is independent of the heating rates.

**Specific heat**

The specific heat measured for sample 1ab and sample 2c is presented in Figure S4(a) and (c). The specific heat of the isostructural non-magnetic $RhCl_3$ is also plotted, which was calculated using two Debye and two Einstein modes according to the parameters reported by S. Widmann *et al* [1]. A scaling factor of 0.92 was applied to the temperature axis of $RhCl_3$. The resulted specific heat of $RhCl_3$ serves as a good approximation to the phonon background ($C_{\mathrm{ph}}$) of $RuCl_3$. Figure S4(c) shows the excess specific heat of sample 1ab: $C_{\mathrm{ex}} = C_{\mathrm{total}} - C_{\mathrm{ph}}$, which is obtained by subtracting the phonon contribution $C_{\mathrm{ph}}$ from the measured total specific heat $C_{\mathrm{total}}$. Figure S4(d) shows the specific heat of sample 1ab measured in various in-plane magnetic fields near the AFM transitions.

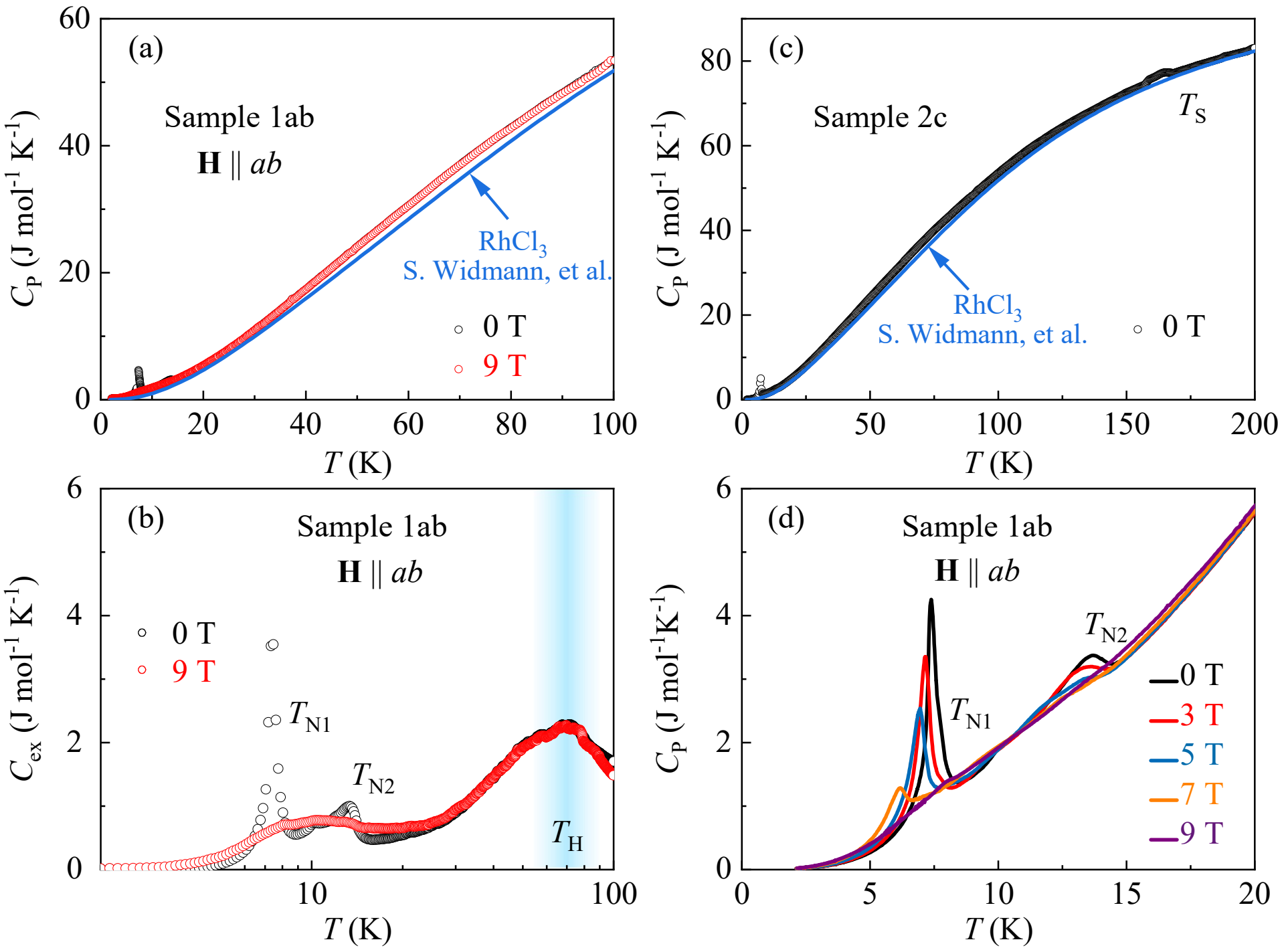

**Figure S4** (a) and (c) Temperature evolution of specific heat ($C_P$) measured for sample 1ab and sample 2c. The data of $RhCl_3$ is evaluated following S. Widmann *et al.*[1], which represents an approximation of the phonon contribution to the specific heat of $RuCl_3$. (b) The excess specific heat of sample 1ab obtained in 0 and 9 T. (c) Low-temperature specific heat of sample 1ab captured in selected in-plane magnetic fields.

## SHG Analysis

Here we analyze the azimuthal angle dependence of SHG signal according to the point group symmetry. The sample lab coordinates are denoted as *x*, *y*, *z*. Under normal incidence with light propagating along the out-of-plane (*z*) direction, nonzero terms of the susceptibility tensor $\chi_{i,j,k}(i,j,k=x,y,z)$ containing in-plane (*x*, *y*) components are expected. The in-plane angle rotation was measured with respect to the crystallographic *a*-axis (*x*) i.e., the $\varphi=0$ direction is perpendicular to the Ru-Ru bonds. The twofold appearance of SHG signal observed both and room temperature and low temperatures is consistence with the monoclinic structure possessing the $C_2$ symmetry. The twofold axis is directed along the *b*-axis (*y* direction). Nonzero components in the susceptibility tensor include three independent elements: $\chi_{xxy}=\chi_{xyx}, \chi_{yxx}, \chi_{yyy}$. The angle evolution of SHG signal is given by:

$$I_{XX}^{2\omega} \propto [\chi_{yyy}\sin^3\varphi + (\chi_{yxx}+2\chi_{xxy})\sin\varphi\cos^2\varphi]^2 I_\omega, \tag{3}$$

and

$$I_{XY}^{2\omega} \propto [\chi_{yxx}\cos^3\varphi + (\chi_{yyy}-2\chi_{xxy})\sin^2\varphi\cos\varphi]^2 I_\omega. \tag{4}$$

As displayed in Figure S5, the data can be traced well with the above $C_2$ treatment.

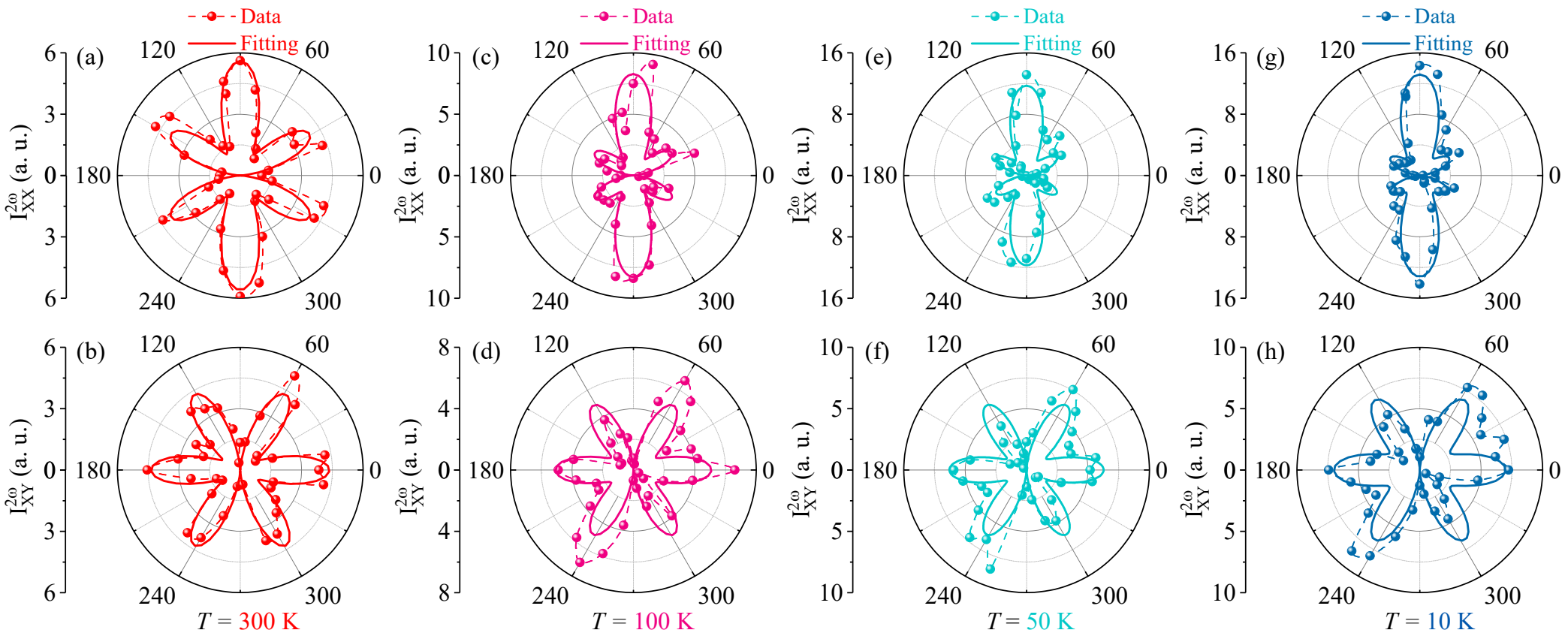

**Figure S5** Theoretical fitting to the azimuthal angle dependence of SHG signal. The $C_2$ symmetry can well describe the SHG results. Solid spheres are experimental data and dashed lines are guide to the eye. Solid lines are theoretical modeling.

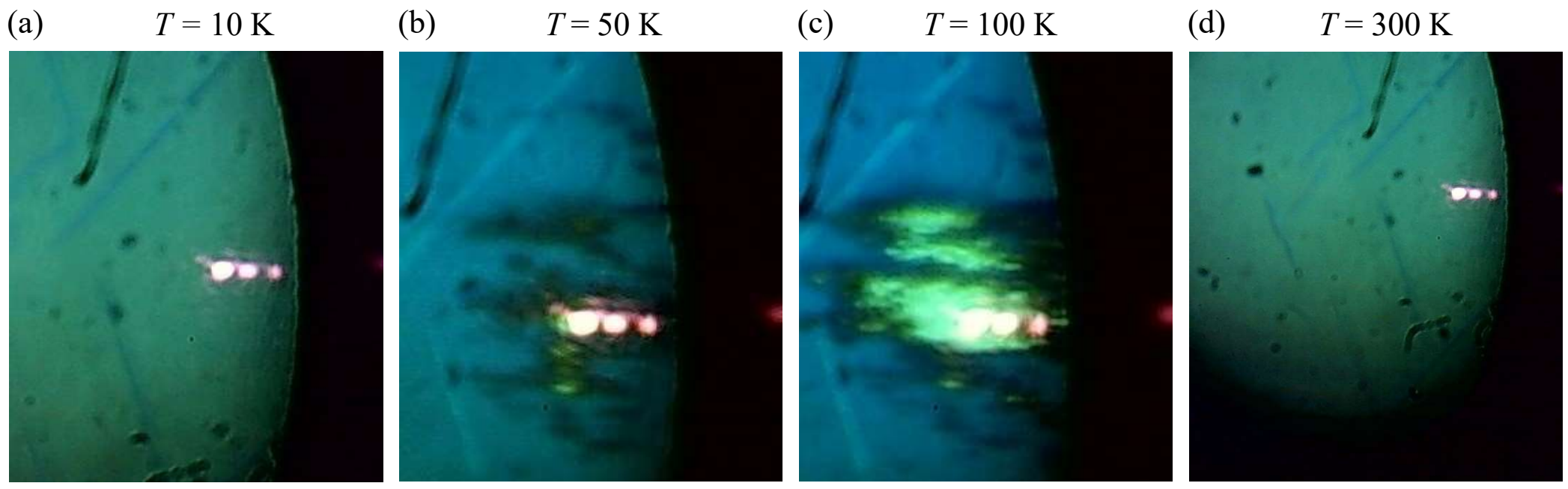

**Figure S6** Optical images of the sample surface taken after the illumination of SHG laser. The excited spot shows greenish color in the crossover temperature interval (50 – 100 K). This phenomenon is absent at other temperatures.

**TEM results**

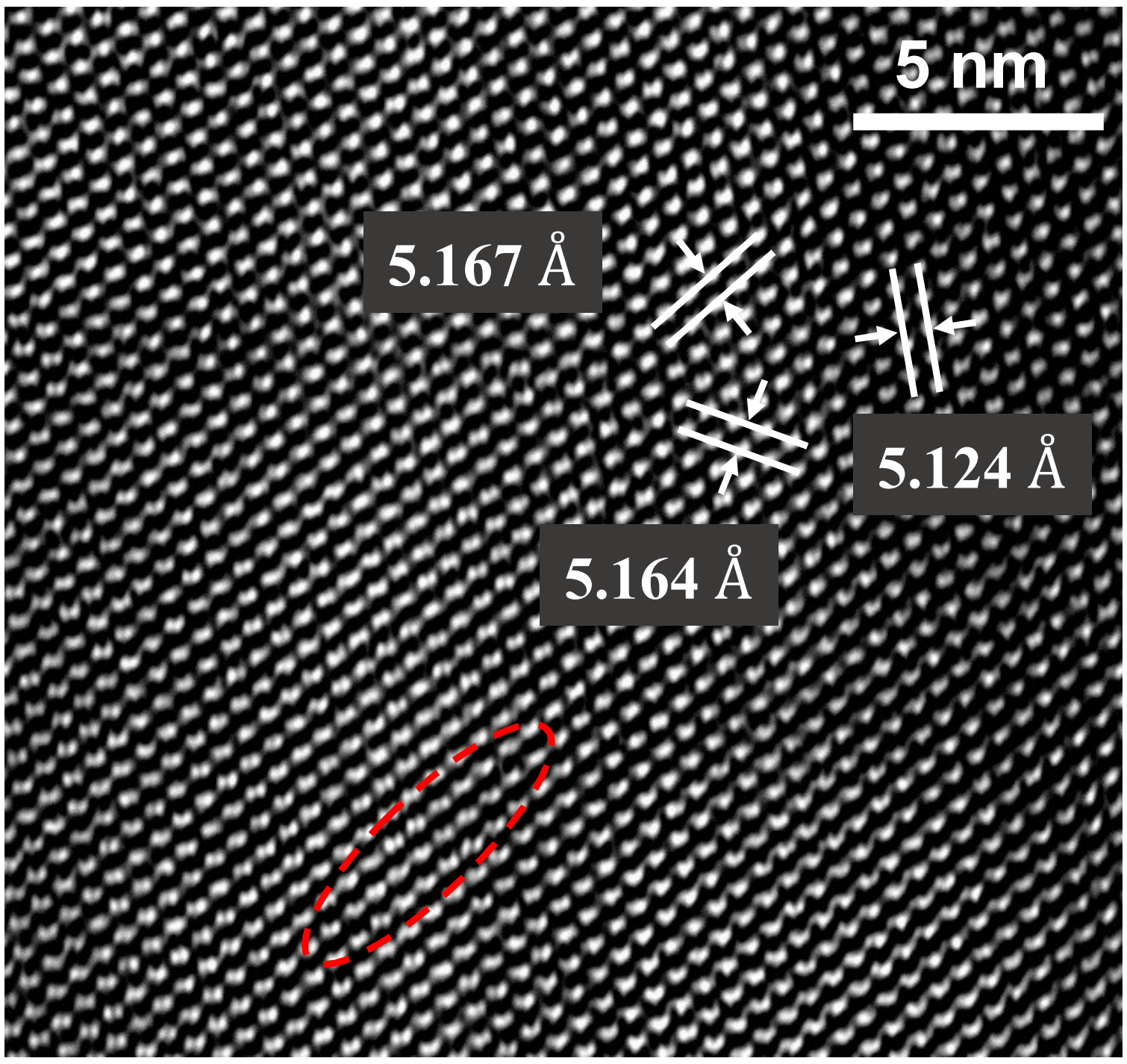

**Figure S7** Typical TEM image of α-$RuCl_3$ taken at room temperature. Bond anisotropies and chain-like structures (highlighted by a red-dashed oval) are seen.

## Electrical control of magnetization

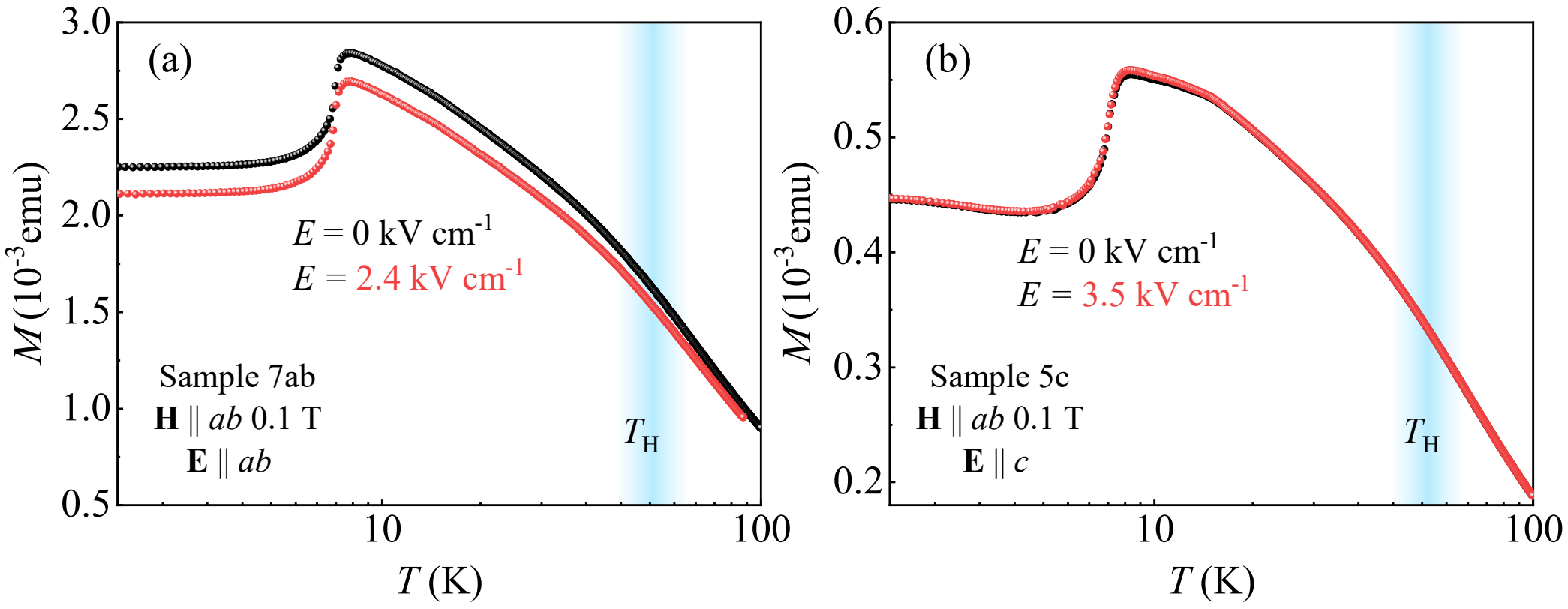

**Figure S8** Electric Field-Tuned Magnetization. The magnetization of α-$RuCl_3$ exhibits moderate changes below $T_H$ in response to in-plane electric fields. Conversely, applying out-of-plane electric fields results in negligible effects.